\def\na{\mbox{$^{22}$\hspace{-0.1em}Na}}            
\def\MeV{\mbox{Me\hspace{-0.1em}V}}                 
\def\Msol{\hbox{M$_{\odot}$}}                       
\def\deg{\hbox{$^\circ$}}                           
\def\funit{\mbox{photons cm$^{-2}$ s$^{-1}$}}       
 
\def\py{\mbox{yr$^{-1}$}}
\def\gray{\mbox{$\gamma$-ray}}                      
\def\apj{ApJ}                                       
\def\mnras{mnras}                                   
\def\aap{AAP}   		             
\def\jrasc{JRASC}   		             

\def   \ni {\noindent}

\def   \ssk {\vskip  5truept}

\def   \bsk {\vskip 15truept}
 
\def   \newpage {\vfill\eject}
\def   \newline {\hfil\break}

\documentstyle[epsfig]{article}
\begin{document}

\hsize 5truein
\vsize 8truein
\font\abstract=cmr8
\font\keywords=cmr8
\font\caption=cmr8
\font\references=cmr8
\font\text=cmr10
\font\affiliation=cmssi10
\font\author=cmss10
\font\mc=cmss8
\font\title=cmssbx10 scaled\magstep2
\font\alcit=cmti7 scaled\magstephalf
\font\alcin=cmr6 
\font\ita=cmti8
\font\mma=cmr8
\def\ref{\par\noindent\hangindent 15pt}
\null


\title{\ni Galactic distribution of 1.275~\MeV\ emission from ONe novae} 
\bsk \bsk

\author{\ni P.~Jean$^{1,2}$, M.~Hernanz$^2$, J.~G\'omez-Gomar$^2$, J.~Jos\'e$^2$}
\bsk

\affiliation{\ni $^1$Centre d'Etude Spatiale des Rayonnements, CNRS/UPS, 9 avenue du colonel Roche, 31028 Toulouse, France, $^2$Institut d'Estudis Espacials de Catalunya (IEEC), Edifici Nexus-201, C/ Gran Capit\`a 2-4, E-08034 Barcelona, Spain}

\bsk
\baselineskip = 12pt
            

\abstract{ABSTRACT \ni Modelisations of galactic 1.275 MeV emission produced by the decay $^{22}$Na have been performed for several frequency-spatial distributions of ONe novae. Recent results of nova rates and their distributions in our Galaxy have been used. These modelisations allow to estimate the lower-limit of the $^{22}$Na mass ejected per ONe novae detectable with the future spectrometer (SPI) of the INTEGRAL observatory as a function of their frequency-spatial distribution in the Galaxy. Calculations using recent estimations of the expected $^{22}$Na mass ejected per ONe nova show that the diffuse galactic 1.275 MeV emission will be difficult to detect with SPI.
}
\bsk
\baselineskip = 12pt
\bsk
\baselineskip = 12pt

\text{\ni 1. INTRODUCTION 
\ssk
\ni    
Observations with several instruments have reported upper-limits on the 1.275~\MeV\ flux from the Galaxy (HEAO-3, SMM) or from individual novae (COMPTEL). Leising et al. (1998) found a 99\%\ confidence limit of 1.2 10$^{-4}$ \funit\ on a steady 1.275~\MeV\ flux from the Galactic center direction. Iyudin et al. (1995), using COMPTEL observations of single novae, estimate a 2$\sigma$ upper-limit of 3 10$^{-5}$ \funit\ for any neon-type novae in the galactic disk, which has been translated into an upper-limit of the ejected \na\ mass of 3.7 10$^{-8}$ \Msol.
The next generation of \gray\ spectrometer should have the sensitivity required for the detection of the 1.275~\MeV\ emission from classical ONe novae. SPI, the future spectrometer of the INTEGRAL observatory will be operationnal in the beginning of 2001. With its high-resolution, the spectrometer is designed for the detection of astrophysical \gray\ lines. SPI will be also able to perform images with an angular resolution of $\approx$3\deg\ by using a coded aperture system. For a more detailed description of the spectrometer see Vedrenne et al. (1998). Using sensitivity estimation and calculations of \na\ yields in ONe novae, Hernanz et al. (1998) estimate that the 1.275~\MeV\ line from a nova could be detected by SPI if its distance is less than $\approx$0.5~kpc. In the presented work, we check the possibility to detect with SPI the cumulative emission from the \na\ ejected by ONe novae. The total Galactic flux at 1.275~\MeV\ depends on the amount of \na\ ejected per outburst and the Galactic ONe nova rate. However, the latter is poorly known because the interstellar extinction prevents us from directly observing in the 
visible more than a small fraction of novae per year. Since 1954, several 
methods have been used to estimate occurence 
\newpage
\ni
rate of novae giving different results. Recently, Shafter (1997) reconciled these differences by recomputing the nova rate with the galactic nova data. He extrapolated the global rate with the observed one accounting for surface brightnesses of the Galaxy and correction factors to take care of any observational incompleteness. He estimated the nova rate to be 35$\pm$11 \py. Hatano et al. (1997) found a similar value (41$\pm$20 \py) using a Monte-Carlo technique with a simple model for the distribution of dust and novae in the Galaxy. Livio \& Truran (1994) estimated the frequency of occurence of ONe novae, in light of observation of abundances in nova ejecta. They estimate a fraction of ONe novae between 11\%\ and 33\%.
We modelize the Galactic emission at 1.275~\MeV\ as a function 
of the ONe novae rate and the mean \na\ yield per outburst. This has been 
done for several spatial distributions of novae in the Galaxy. The observation time necessary for a detection with SPI of \na\ emission is estimated as a function of the mean mass of \na\ ejected per nova.

\bsk
\ni 2. METHOD 
\ssk
\ni The method consists of: (1) a simulation of a set of ONe novae that is
representative of what could be the galactic novae distribution at a given
time. (2) An analysis of the distribution of the 1.275~\MeV\  emission and a check of whether SPI will be able to detect it. And, at least, (3) an estimation of the frequency of detection by analysing a large number of Galaxy-tests.

\ssk
\ni 2.1 Modelisation of galactic 1.275 \MeV\ emission
\ssk
\ni 
The galactic distribution of the 1.275~\MeV\ emission from ONe novae is calculated with a Monte-Carlo simulation. The position in Galactocentric coordinates and the age of ONe novae are chosen randomly according to the appropriate distributions. The number of simulated ONe novae depends on their frequency and the total period during which ejected \na\ is an effective emitter. As proposed by Higdon \& Fowler (1987, hereafter HF87), it has been assumed that the emissivity of novae older than 5 times the \na\ mean life contributes negligibly to the diffuse 1.275~\MeV\ emission. The \na\ mass ejected per nova and the rate of novae are parameters of the modelisation.
We have selected four models of distribution of novae in the Galaxy that differ significantly each other. The first of them 
is described in HF87. They assumed that the novae are distributed like stars. The second model has been used by Hatano et al. (1997) to estimate the spatial distribution and the rate of Galactic novae. It is based on a model of the distribution of SNIa by Dawson \& Johnson (1994, hereafter DJ94). Since SNIa are probably the result of the thermonuclear explosion of a mass-accreting white dwarf, their distribution should be close to the novae one. The third model is derived from the galactic survey of the Spacelab InfraRed Telescope that provides a reliable tracer of the distribution of G and K giant stars (i.e. old population stars - see Kent, Dame \& Fazio, 1991, hereafter KDF91). The last model is taken from Van der Kruit (1990, hereafter VdK90). It has been used by Shafter (1997) to estimate the nova rate in our Galaxy. He assumed that the nova distribution follows the brightness profile of our Galaxy. According to recent results (see section 1), we did calculations for ONe nova rate ranging from 2~\py\ to 18~\py.
\ssk
\ni 2.2 Observation of the galactic 1.275 \MeV\ emission by SPI
\ssk
\ni
Jean et al. (1997) have estimated the narrow line sensitivity of SPI for an on-axis point source by computing instrumental background with detailed physics Monte-Carlo simulations. However, the 1.275~\MeV\ line is 20~keV width (see Hernanz et al., 1998) and this sensitivity become 2.2 10$^{-5}$ \funit\ (3$\sigma$, 10$^6$ s observation time). Moreover, the emission is diffuse and the spectrometer will provide (1) a distribution of intensity in pixels and (2) a total flux in the field-of-view. The both cases have been investigated for the analysis of Galaxy-tests. In the case 1, distribution of the intensity has been calculated with 3\deg\ by 3\deg\ size pixels (SPI angular resolution). A rough estimation of the significance, for a detection of at least one excess in the distribution of the intensity, is derived by computing the probability that the observed distribution is due to background fluctuations. The analysis in the case 2 is similar to an on/off pointing method analysis. The sensitivity of such a mode of detection is 3.1 10$^{-5}$ \funit.
\ssk
\ni 2.3 Probability to detect the 1.275 \MeV\ line with SPI
\ssk
\ni 
Several Monte-Carlo simulations have been done for a given galactic ONe-novae frequency-spatial distribution and a value of the \na\ yield per nova. For a large number of Galaxy-tests, the probability for a detection of the cumulative emission has been estimated by calculating the fraction of time the observation of the simulated flux give a significance larger than 3$\sigma$. The average observation time  needed to have 90\%\ of chance for a detection of this emission has been derived.

\bsk
\ni 3. RESULTS
\ssk
\ni
Figure 1 shows the observation time needed to have 90\%\ of chance that SPI detect the 1.275 \MeV\ line (case 2) from the GC region as a fonction of the mean \na\ mass ejected per ONe novae. These results are displayed for the 4 spatial distributions. The time needed for a detection of the cumulative emission for models KDF91 and HF87 are lower than those of VdK90 and DJ94. The Galactocentric scale radius of the exponential disk for the former models are 3~kpc and 3.5~kpc respectively whereas it is higher (5~kpc) for the model VdK90 and DJ94. Since the novae are closer to the GC in the KDF91 and HF87 models, the 1.275~\MeV\ flux is larger and more compact. The mean simulated 1.275~\MeV\ flux from the 12.5\deg\ around the GC, normalized to the \na\ yield per outburst and the rate of ONe novae (in \funit \Msol$^{-1}$ yr), are 78.0$\pm$2.4, 56.8$\pm$3.5, 105.1$\pm$2.9 and 56.4$\pm$1.4 for the HF87, VdK90, KDF91 and DJ94 model respectively. SPI will need 10 times more observation time for a detection of an excess in the pixel distribution. 

\begin{figure}
 \centerline{\mbox{\psfig{file=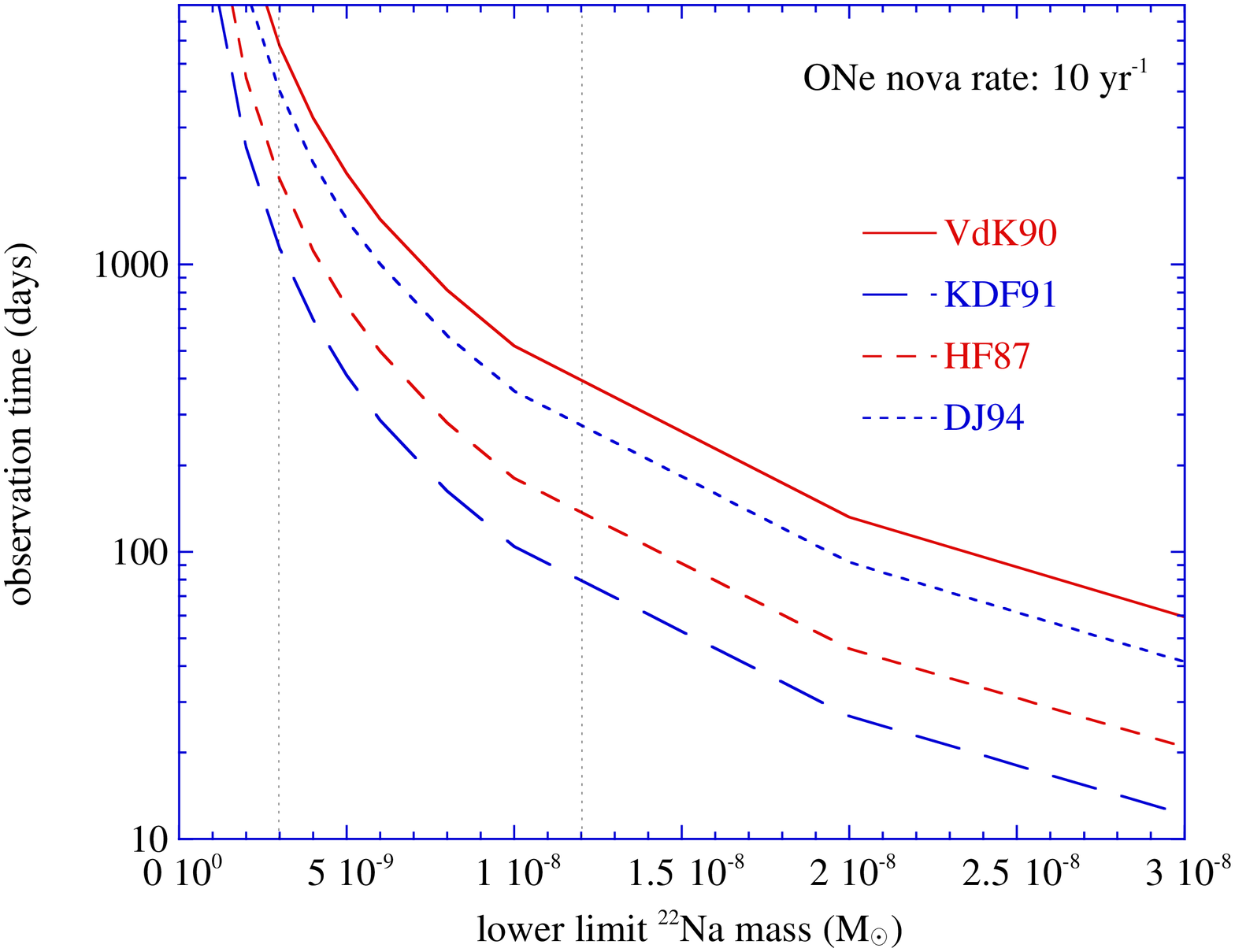,height=5.0cm,width=8.cm}}}
 \caption{Figure 1: SPI observation time needed for a detection of the 1.275 \MeV\ emission of the GC region as a function of the \na\ yield per nova and for 4 galactic distributions.}
\end{figure} 

\bsk
\ni 4. CONCLUSION 
\ssk
\ni 
According to Hernanz et al. (1998), the \na\ average yield could be between
3~10$^{-9}$ and 1.2~10$^{-8}$ \Msol\ per nova. There is few chance to detect the galactic diffuse 1.275~\MeV\ emission with SPI with $\approx$10 days of observations. However, a 80~days of observation of the GC region could already give constraints for the mean \na\ yield per ONe novae and for their distribution in the Galaxy. Therefore, \gray\ observation of novae would provide information not only on their eruption mechanisms and the nucleosynthesis processes involved in their explosion but also on their distribution in the Galaxy.
}

\bsk
\baselineskip = 12pt
{\abstract \ni ACKNOWLEDGMENTS
Research partially supported by the training and Mobility Researchers Programme. Access to large installation, under contract ERBFMGECT050062 - Access to supercomputing facilities for european researchers established between the European Community and Community and CESCA-CEPBA.
}
\ssk
{\references \ni REFERENCES
\ssk
\ref {Dawson}, P.~C., \& {Johnson}, R.~G. 1994, \jrasc, 88, 369
\ref {Hatano}, K., {Branch}, D., {Fisher}, A., \& {Starrfield}, S. 1997, \mnras, 290, 113--118
\ref {Hernanz}, M., {Gomez-Gomar}, J., {Jose}, J., \& {Isern}, J. 1998, in these procedings
\ref {Higdon}, J.~C., \& {Fowler}, W.~A. 1987, \apj, 317, 710--716
\ref {Iyudin}, A.~F., et al., 1995, \aap, 300, 422
\ref {Jean}, P., et al., 1997, Proc. of the 2$^{nd}$ INTEGRAL Workshop, ESA publication division, 382, 635
\ref {Kent}, S.~M., {Dame}, T.~M., \& {Fazio}, G. 1991, \apj, 378, 131-138
\ref {Leising}, M.~D., {Share}, G.~H., {Chupp}, E.~L., \& {Kanbach}, G. 1988, 
\apj, 328, 755--762
\ref {Livio}, M., \& {Truran}, J.~W. 1994, \apj, 425, 797--801
\ref {Shafter}, A.~W. 1997, \apj, 487, 226
\ref {Van der Kruit}, P. 1990, in The Milky Way as a Galaxy, ed. R. Buser \&\ 
I.R. King (Mill Valley: University Science Books), p. 331
\ref {Vedrenne}, G. et al., 1998, in these procedings
}

\end{document}